\pdfoutput=1
\newlength{\absize}
\documentclass[12pt]{article}
\usepackage{graphicx,dsfont,subfig,amsmath,amssymb,setspace,sectsty,hyperref}
\hypersetup{linktocpage}
\renewcommand{\baselinestretch}{1.5}

\sectionfont{\large}
\numberwithin{equation}{section}
\begin{document}
\thispagestyle{empty}
\pagestyle{empty}
\renewcommand{\thefootnote}{\fnsymbol{footnote}}
\newcommand{\starttext}{\newpage\normalsize
\pagestyle{plain}
\setlength{\baselineskip}{3.5ex}\par
\setcounter{footnote}{0}
\renewcommand{\thefootnote}{\arabic{footnote}}}
\newcommand{\preprint}[1]{\begin{flushright}
\setlength{\baselineskip}{3ex}#1\end{flushright}}
\renewcommand{\title}[1]{\begin{center}\Large\bf
#1\end{center}\par}
\renewcommand{\author}[1]{\vspace{2ex}{\normalsize\begin{center}
\setlength{\baselineskip}{3.25ex}#1\par\end{center}}}
\renewcommand{\thanks}[1]{\footnote{#1}}
\renewcommand{\abstract}[1]{\vspace{2ex}\normalsize\begin{center}
\centerline{\bf Abstract}\par\vspace{2ex}\parbox{\absize}{#1
\setlength{\baselineskip}{3.25ex}\par}
\end{center}}
\setcounter{bottomnumber}{2}
\setcounter{topnumber}{3}
\setcounter{totalnumber}{4}
\renewcommand{\bottomfraction}{1}
\renewcommand{\topfraction}{1}
\renewcommand{\textfraction}{0}
\def\draft{
\renewcommand{\label}[1]{{\quad[\sf ##1]}}
\renewcommand{\ref}[1]{{[\sf ##1]}}
\renewenvironment{equation}{$$}{$$}
\renewenvironment{thebibliography}{\section*{References}}{}
\renewcommand{\cite}[1]{{\sf[##1]}}
\renewcommand{\bibitem}[1]{\par\noindent{\sf[##1]}}
}
\def\theequation{\thesection.\arabic{equation}}
\preprint{}
\newcommand{\be}{\begin{equation}}
\newcommand{\ee}{\end{equation}}
\newcommand{\ba}{\begin{eqnarray}}
\newcommand{\ea}{\end{eqnarray}}
\newcommand{\bas}{\begin{eqnarray*}}
\newcommand{\eas}{\end{eqnarray*}}
\newcommand{\bc}{\begin{center}}
\newcommand{\ec}{\end{center}}
\newcommand{\nn}{\nonumber}
\newcommand{\comment}[1]{}
\newcommand{\csch}{\mathop{\rm csch\,}}
\newcommand{\sech}{\mathop{\rm sech\,}}

\title{Linear Confinement of Quarks from Supergravity}
\author{Girma Hailu\thanks{hailu@physics.harvard.edu}\\
\vspace{2ex}
\it{Jefferson Physical Laboratory\\
Harvard University\\
Cambridge, MA 02138} }

\abstract{
A supergravity background that produces linear confinement of quarks in four dimensions is presented.
}

\starttext
\tableofcontents

\section{\label{sec:intro}Introduction}

Experimental data show mass spectra of mesons and baryons in which the spins of the particles are approximately linear
in the mass squares with nearly universal slope of $\alpha'\approx 1\,GeV^{-2}$ \cite{Nakamura:2010zzi}. Determining the mass spectra  using the theory of quantum chromodynamics (QCD) is a challenging problem, because of the nonperturbative nature of confinement.
Over four decades ago, the mass spectra were related to linear Regge trajectories in scattering amplitudes and it was observed that the universality of the slope of the trajectories suggested a dual string model of mesons and baryons with quarks and antiquarks at the ends of the string  \cite{Veneziano:1968yb}. The strings in the early dual models lived in the same four-dimensional (4D) spacetime as QCD, and it was realized that the string amplitudes had exponential decay at large transverse momenta inconsistent with data which show power-law behavior. Approaches in which the strings propagate in five or higher dimensions with the string tension running in the extra space were suggested later \cite{Polyakov:1997tj}. Furthermore, a consistent relativistic quantum theory of superstrings lives in ten dimensions (10D).
In a parallel development, it was learned that the actual perturbative expansion parameter of QCD, generalized to $N$ colors, was the 't Hooft coupling $\frac{g_{4}^2N}{4\pi}$, where $g_{4}$ is the Yang-Mills coupling, and that QCD might be dual to a worldsheet theory of strings in the large $N$ limit with the 't Hooft coupling kept fixed \cite{'tHooft:1973jz}.

The first concrete example of gauge/string (or gauge/gravity) correspondence was obtained in \cite{Maldacena:1998re}, where it was found that type IIB string theory on $AdS_5\times S^5$ was dual to $\mathcal{N}=4$  $SU(N)$ conformal field theory that lives on a stack of $N$ parallel D3-branes.
Perturbative gauge theory description is appropriate for a 't Hooft coupling that is smaller than $1$, while the gravity description is appropriate for a 't Hooft coupling that is much bigger than $1$. It was also learned that the potential energy of a quark-antiquark pair in 4D on backgrounds involving $AdS_5$ metric goes as $\frac{1}{L}$ \cite{Maldacena:1998im}, where $L$ is the length of the string, which does not confine, consistent with conformal field theory.

We proposed and argued in \cite{Hailu:PN1} a correspondence between type IIB string theory with $N$ D7-branes on $\mathds{R}^{1,3}\times \frac{\mathds{C}^1}{Z_2}\times \frac{\mathds{T}^2}{Z_2}\times \frac{\mathds{T}^2}{Z_2}$ and pure $\mathcal{N}=1$ $SU(N)$ gauge theory in 4D and showed that the background reproduces the renormalization group flow and the pattern of chiral symmetry breaking of the gauge theory and discussed that it confines.

In this letter, it is shown that the background in \cite{Hailu:PN1} produces linear confinement of quarks in 4D with the energy of a quark-antiquark pair increasing linearly with the length of the string in between.

\section{Background geometry\label{backm-1}}

The background metric is briefly summarized in this section, see \cite{Hailu:PN1} for details. The extra dimensional space is parameterized by one radial coordinate $r$ with range $1\le r\le \,e^{\frac{2\pi}{3g_sN}}$, where $r$ is measured in units of $r_s=\sqrt{\alpha'}$ here, and five angles $\psi$, $\varphi_1$, $\varphi_2$, $\varphi_3$, and $\varphi_4$, each having the range between 0 and $2\pi$; $r$ and $\psi$ are coordinates on $\frac{\mathds{C}^1}{Z_2}\sim \mathds{R}^1\times\mathds{S}^1$ and the remaining angles are coordinates on $\frac{\mathds{T}^2}{Z_2}\times \frac{\mathds{T}^2}{Z_2}$. We call $r=1$ the infrared (IR) boundary and $r=e^{\frac{2\pi}{3g_sN}}$ the ultraviolet (UV) boundary.
There are $N$ D7-branes wrapped over $\frac{\mathds{T}^2}{Z_2}\times \frac{\mathds{T}^2}{Z_2}$ at $r=1$.
The radial coordinate $r$ is mapped to the scale $\Lambda$ of the gauge theory as  $\Lambda\sim r$. In particular, the scale at which the gauge coupling formally diverges is mapped to $r=1$. The coordinate $\psi$ is mapped to the Yang-Mills angle $\Theta$ in the gauge theory as $\Theta=N\psi$.
The metric, with $x^\mu$ and $r$ all measured in units of $r_s=\sqrt{\alpha'}$, is
\ba
&ds_{10}^2=r_s^2\,\Bigl(\cosh u\,dx_{1,3}^2
+ \sech u\,(9\,\frac{dr^2}{r^2}+d\psi^2&\nn\\&+d\varphi_1^2+d\varphi_2^2
+d\varphi_3^2+d\varphi_4^2)\Bigr),\label{metric-3}&
\ea
where
\be
\sech u= \sqrt{1-(\frac{g_sN}{2\pi}\rho)^2},\quad \rho\equiv 3 \ln {r},\label{h1}
\ee
and $dx_{1,3}^2$ is the metric on flat 4D spacetime $\mathds{R}^{1,3}$ with coordinates $x^\mu$, $\mu=0,1,2,3$. We use uppercase letter $X$ with uppercase indices $M,N,\cdots$ to denote all coordinates of the 10D spacetime.
The geometry is compact. The curvature is smallest in the IR region near $r=1$ where the gauge theory is strongly coupled and a dual gravity description is useful. The internal space normal to the D7-branes is $\mathds{S}^1$ at the IR boundary whose radius is set by the nonperturbative scale of the gauge theory and spacetime is 4D at the UV boundary. The UV boundary provides a convenient setting for putting quarks and antiquarks and branes that serve as sources of flavor symmetry.

\section{Linear confinement\label{sec-conf}}

The warp factor in the metric given by (\ref{metric-3}) is nonzero everywhere and increases with increasing $r$.
One measure of confinement is the area bounded by a Wilson loop \cite{Wilson:1974sk}. It follows from the metric that the area enclosed by a Wilson loop in 4D at some $r$ between the IR and the UV boundaries is minimized for a surface stretched and bent toward the IR boundary.

Let us find out the specific features by studying the energy and the configuration of the string in a quark-antiquark pair.
Consider a quark and an antiquark in 4D located at $(x^1,x^2,x^3)=(\pm \frac{L}{2},0,0)$ at the UV boundary connected by an oriented open string.
We may view the location of the quark and the antiquark to be a probe D3-brane. The string worldsheet action is given by
\be
S_2=\frac{1}{2\pi\alpha'}\int d\sigma^0d\sigma^1\, \sqrt{\det G_{MN}\partial_{a}X^M \partial_{b}X^N}\,,\label{S2-1}
\ee
where $(\sigma^0,\,\sigma^1)$ parameterize the coordinates of the two-dimensional string worldsheet, $a$ and $b$ denote the indices, and $G_{MN}$ is the metric on the 10D space.
To find the configuration of the string, we take spacetime to be Euclidean and choose $\sigma^0=x^0$, $\sigma^1=x^1$ (we write $x^1$ as $x$ in the following), and $\rho=\rho(x)$.
For the evolution of the string between time $x^0_a$ and $x^{0}_b$, the Wilson loop is a rectangle on the UV boundary with corners at $(x^0,\,x)=(x^0_a,\,\pm\frac{L}{2})$ and $(x^0_b,\,\pm\frac{L}{2})$. The corresponding string worldsheet is an area bounded by the Wilson loop.

We want a static configuration of the string at some time $X^0\to\infty$, and the action reduces to
\be
S_2=\frac{X^0}{2\pi}\int dx\, \sqrt{\cosh^2u+(\frac{d\rho}{dx})^2}\,.\label{S2-2}
\ee
To minimize the area, we extremize the action using the Euler-Lagrange equation, and obtain
\be
-\frac{X^0}{2\pi}\frac{\cosh^2u}{\sqrt{\cosh^2u+(\frac{d\rho}{dx})^2}}
=\mathrm{constant}.\label{const-1}
\ee
Because the quark and the antiquark are put symmetrically at $x=\pm \frac{L}{2}$, we have $\rho(-x)=\rho(x)$, $(\frac{d\rho}{dx})_{\rho=0}=0$, the constant in (\ref{const-1}) is $-\frac{X^0}{2\pi}$, and in the $x>0$ region,
\be
\frac{dx}{d\rho}={\sech u \,\csch u}=\frac{2\pi}{g_sN}(\frac{1}{\rho}-(\frac{g_sN}{2\pi})^2\rho).
\label{xt-1aa}
\ee
Using (\ref{xt-1aa}) in (\ref{S2-2}) and evaluating the integral for the whole range $-\frac{L}{2}\le x \le \frac{L}{2}$, the action is expressed in terms of the length of the string in 4D,
\be
S_2=2\left(\frac{X^0}{2\pi}(x-\frac{1}{2}{\sech}^2 u)|_{x=0}^{L/2}\right)=\frac{X^0}{2\pi}(L+1).\label{S2-3}
\ee

Thus the energy of the string is
\be
E=\frac{1}{2\pi}\,L \label{S2-4}
\ee
for $L>>1$, which increases linearly with the length of the string in 4D. The string tension equals $\frac{1}{2\pi}$ in string units.

Next we find the explicit relation between $x$ and $\rho$ and analyze the configuration of the string between the quark and the antiquark. First let us see the basic features qualitatively. We notice from (\ref{xt-1aa}) that $\frac{d\rho}{dx}=0$ at  $\rho=0$. Therefore, $x$ increases and the string is stretched to infinity at $\rho=0$. As $\rho$ increases and for $\rho\approx 0$, we have $\frac{d\rho}{dx}\sim \frac{g_sN}{2\pi} \rho$ and $x\sim\frac{2\pi}{g_sN}\ln \rho$. When $\rho$ gets closer to the UV boundary, where $\rho=\frac{2\pi}{g_sN}$, we have $\frac{dx}{d\rho}\to 0$ and $x$ barely changes in the transverse $\rho$ direction. At the UV boundary, $\frac{dx}{d\rho}= 0$ and the two sides of the string end on the quark and the antiquark at normal angle to the 4D space.

More quantitatively, the solution to (\ref{xt-1aa}) with the boundary condition that $x= \frac{L}{2}$ at $\rho=\frac{2\pi}{g_sN}$ is
\be
x=\frac{L}{2}+\frac{2\pi}{g_sN}\left(\ln (\frac{g_sN}{2\pi}\rho)+\frac{1}{2}(1- (\frac{g_sN}{2\pi}\rho)^2)\right).
\label{xt-1}
\ee
Thus $x$ has logarithmic divergence in $\rho$ at $\rho=0$, since the string gets stretched to $x=\infty$ at $\rho=0$. We regularize the divergence by introducing a cutoff and limiting the range of $\rho$ to $\rho_0\le\rho\le \frac{2\pi}{g_sN}$, and thereby making the length (and the energy) of the string finite, with $\rho_0>0$ and $x=0$ at $\rho=\rho_0$. The values of $\rho_0$ and ${L}$ are related by
\be
\frac{g_sN}{2\pi}\frac{L}{2}+\frac{1}{2}=\ln (\frac{2\pi}{g_sN}\frac{1}{\rho_0})+\frac{1}{2}(\frac{g_sN}{2\pi}\rho_0)^2
\label{rhoL-2}
\ee
and (\ref{xt-1}) can be rewritten with $L$ expressed in terms of $\rho_0$ as
\be
x=\frac{2\pi}{g_sN}\left(\ln (\frac{\rho}{\rho_0})-\frac{1}{2} (\frac{g_sN}{2\pi})^2(\rho^2-\rho_0^2)\right).
\label{xt-2}
\ee
Notice that $\rho_0\to 0$ as $L\to \infty$ and $\rho_0\to \frac{2\pi}{g_sN}$ as $L\to 0$. Because we are studying low-energy confining physics, we need to take large $L$ and  $\rho_0$ is close to $0$.

Thus the area bounded by the Wilson loop in 4D at the UV boundary is minimized for a string between the quark and the antiquark that is stretched toward the IR boundary. In fact, for $L>>1$ we may view the string in between as two open strings coming from the quark and the antiquark and ending on the D7-branes which give color to the quark and the antiquark that interact through their color charges. Because $L>>1$, spacetime is effectively 4D for an observer who measures the interaction between the quark and the antiquark.

\begin{figure}[t]
\begin{center}
\includegraphics[width=1.85in]{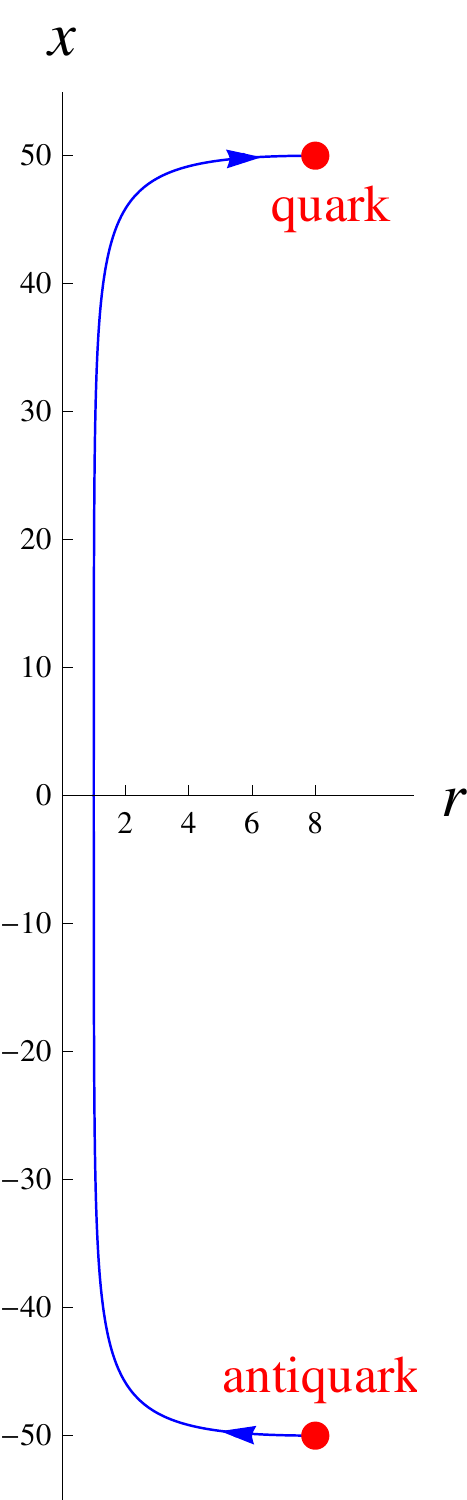}
\end{center}
\caption{Static configuration of the string between the quark and the antiquark for $g_sN=1$ and $L=100$.}
\label{fig:conf1}
\end{figure}

Figure \ref{fig:conf1} shows a configuration of the stretched string, plotted using the analytic expression given by (\ref{xt-2}) with $\rho=3 \ln r$ and $e^{\frac{\rho_0}{3}}\le r\le e^{\frac{2\pi}{3\check{}}}$ for $g_sN=1$ and $L=100$ which corresponds to $\rho_0=1.3\times 10^{-3}$; $g_sN$ is the 't Hooft coupling at the UV boundary. The 't Hooft coupling at $\rho$ between $\rho_0$ and $\frac{2\pi}{g_sN}$ is $\frac{2\pi}{\rho}$ which is large in the IR region near $\rho=\rho_0$.

Thus the quark and the antiquark are confined and the gravity theory provides a realization of linear confinement in 4D with the energy increasing linearly with the length of the string in the IR.

\section{Conclusions\label{concl}}

The supergravity background which we argued in \cite{Hailu:PN1} to correspond to pure $\mathcal{N}=1$ supersymmetric $SU(N)$ gauge theory in 4D in the IR robustly produces linear confinement of quarks. It means $\mathcal{N}=1$ $SU(N)$ gauge theory has similar Regge-type mass trajectories as in QCD.
String theory had its origin over four decades ago in attempts to obtain Regge-type mass spectra using dual string models.
Linear confinement is produced within a supergravity background of consistent relativistic quantum theory of type IIB strings in this letter for the first time.

The background can be used for modeling hadrons and their masses in $\mathcal{N}=1$ $SU(N)$ gauge theory by putting quarks and antiquarks on the 4D space at the UV boundary with the wrapped D7-branes at the IR boundary serving as sources of color, as described in this letter.

Varying scales of string tension, which give varying slopes of mass trajectories similar to what data show in QCD, may be obtained with different string lengths in 4D which give different values of the IR regularization parameter $\rho_0$.

The UV boundary can be used not only as a setting where quarks and probe branes which provide sources of flavor symmetry could reside but also as a UV cutoff to the gravity theory, with appropriate choice of $g_sN$, beyond which perturbative gauge theory description is appropriate.

Because the gravity theory contains crucial features of the non-supersymmetric strong nuclear interactions, gauge coupling running, chiral symmetry breaking, and linear confinement, it might also be useful for exploring nonperturbative phenomena in QCD.


\end{document}